\documentclass[12pt]{article}
\usepackage{amsfonts}
\usepackage[dvips]{color}
\usepackage{psfrag}
\usepackage{feynmp}
\usepackage[pdftex]{graphicx}
\usepackage{braket}
\usepackage{bm}
\usepackage{subfigure}

\usepackage{ulem}
\usepackage{comment}
\includecomment{pdffig}

\usepackage[height=8.5in,width=6.4in]{geometry} 
\usepackage{xparse}
\usepackage{xcolor}
\usepackage{tikz}
\usepackage{cite}
\usepackage[linktocpage]{hyperref}
\usepackage{amssymb, mathtools}
\usepackage[vcentermath]{youngtab}

\usetikzlibrary{cd}

\allowdisplaybreaks[1]

\Yboxdim{5pt}

\makeatletter
    
    \@addtoreset{equation}{section}
  \makeatother

\def\rem#1{}

\renewcommand{\title}[1]{\vbox{\center\LARGE{#1}}\vspace{5mm}}
\renewcommand{\author}[1]{\vbox{\center\large#1}\vspace{5mm}}

\numberwithin{equation}{section}

\DeclareFontShape{OT1}{cmr}{mx}{n}%
    {<->cmr10}{}

\renewcommand{\hat}{\widehat}

\allowdisplaybreaks

\begin{document}
\bibliographystyle{utphys}

 \begin{titlepage}
 \begin{center}
 \vspace{5mm}
 \hfill {\tt 
 NITEP 284
 }\\
 \vspace{20mm}

 \title{
 \LARGE  
  3d SUSY enhancement with non-trivial Coulomb branch via 4d $\mathcal{N}=2$ SCFT
}
 \vspace{7mm}

 Ryo Hamachika,$^{a}$ Takahiro Nishinaka,$^{a,b,c}$ Shou Tanigawa$^{a}$ and 
 Yutaka Yoshida$^{d,e}$

 \vspace{6mm}

 \vspace{3mm}
 $^a${\small {\it Department of Physics, Graduate School of Science
 Osaka Metropolitan University, Osaka 558-8585, Japan}} \\
 $^b${\small {\it Nambu Yoichiro Institute of Theoretical and Experimental Physics (NITEP)
 Osaka Metropolitan University, Osaka 558-8585, Japan}} \\
 $^c${\small {\it  Osaka Central Advanced Mathematical Institute (OCAMI)
 Osaka Metropolitan University, Osaka 558-8585, Japan}} \\
 $^d${\small {\it Department of Current Legal Study, Faculty of Law,  Meiji Gakuin University, 1-2-37 
 Shirokanedai, Minato-ku, Tokyo 108-8636, Japan}} \\
 $^e${\small {\it Institute for Mathematical Informatics, Meiji Gakuin University
 1518 Kamikurata-cho, Totsuka-ku, Yokohama, Kanagawa 244-8539, Japan}}

 \end{center}

 \vspace{7mm}
 \abstract{
In recent years, a variety of three-dimensional (3d) $\mathcal{N}=2$ Chern--Simons (CS) matter theories have been constructed that flow to 3d $\mathcal{N}=4$ superconformal field theories (SCFTs) obtained from $R$-twisted reductions of four-dimensional (4d) $\mathcal{N}=2$ SCFTs. In all previously studied examples, the resulting 3d $\mathcal{N}=4$ SCFTs have trivial Coulomb branches.
In this paper, we construct a 3d $\mathcal{N}=2$ CS matter theory that flows to the $R$-twisted reduction of the $(A_2,D_4)$ Argyres--Douglas theory. This gives the first example in this class whose infrared 3d $\mathcal{N}=4$ SCFT has a non-trivial Coulomb branch, which we argue is given by the orbifold $\mathbb{C}^2/\mathbb{Z}_2$, while its Higgs branch is trivial. We further identify the 4d $U(1)$ R-charge as a mixing of a 3d R-charge with an emergent 3d Coulomb branch flavor charge that has no 4d origin.
 }
 \vfill

 \end{titlepage}

\tableofcontents

\section{Introduction}

It has recently been discovered that the R-twisted circle
compactification of 4d $\mathcal{N}=2$ superconformal field theories
(SCFTs) on a cigar geometry gives rise to topologically twisted 3d
$\mathcal{N}=4$ SCFTs on a half space with a holomorphic boundary
condition \cite{Dedushenko:2023cvd, Dedushenko:2018bpp, ArabiArdehali:2024ysy, Gaiotto:2024ioj,
ArabiArdehali:2024vli, Costello:2018fnz}. An intriguing feature
of these 3d $\mathcal{N}=4$ SCFTs is that they can be realized as IR
fixed points of 3d $\mathcal{N}=2$ Chern-Simons (CS) matter theories
with appropriate monopole superpotential deformations, where a
supersymmetry enhancement from $\mathcal{N}=2$ to $\mathcal{N}=4$ occurs
in the infrared \cite{Gang:2018huc, Gang:2021hrd, Gang:2023rei}.
Given these recent discoveries, it is an important problem to identify a 3d $\mathcal{N}=2$ CS matter theory whose
IR fixed point describes the R-twisted circle compactification of a given 4d
$\mathcal{N}=2$ SCFT. Indeed, many 3d $\mathcal{N}=2$ CS matter theories
have been identified in the literature as
corresponding to various 4d $\mathcal{N}=2$ SCFTs
\cite{Dedushenko:2023cvd,
Gang:2023rei, Gang:2023ggt, Ferrari:2023fez, Baek:2024tuo,
Creutzig:2024ljv, ArabiArdehali:2024ysy, Gaiotto:2024ioj,
ArabiArdehali:2024vli, Gang:2024loa, Kim:2024dxu, Go:2025ixu, Kucharski:2025lcr,
Kim:2025klh, Kim:2025rog, Nishinaka:2025ytu}.

However, as far as the authors are aware, all the 4d $\mathcal{N}=2$
SCFTs studied so far in this context are those without a Coulomb branch
generator of integer
$U(1)_r$ charge. In that case, all the 4d Coulomb branch operators are
removed from the spectrum by the R-twisted circle reduction, leading to
a trivial Coulomb branch in three dimensions. This drastically simplifies the correspondence
between the 4d and 3d quantities. For instance, since the 3d Coulomb
branch is trivial, the 3d $\mathcal{N}=4$ flavor symmetry, if it exists, always acts on
the Higgs branch. Since the 4d and 3d Higgs branches are expected to be
identical, one can then easily identify the 3d $\mathcal{N}=4$ flavor
symmetry from the flavor symmetry of the original 4d $\mathcal{N}=2$
SCFT. If the 3d Coulomb branch is non-trivial, we do not have such
simplicity, since a flavor symmetry acting on a 3d Coulomb branch has no
4d counterpart and therefore is an accidental symmetry in three dimensions.\footnote{One can show that 4d Coulomb branch operators are
always neutral under any 4d $\mathcal{N}=2$ flavor symmetry
\cite{Buican:2013ica, Buican:2014qla}.}

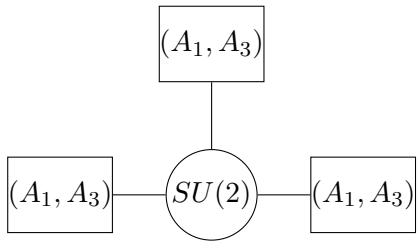
\begin{figure}
\centering
 \begin{tikzpicture}[vec/.style={thick,-latex,draw=black},gauge/.style={circle,draw=black,inner sep=0pt,minimum size=12mm},flavor/.style={rectangle,draw=black,inner sep=0pt,minimum size=10mm},auto]
  \node[gauge] (1) at (0,0) {\small $SU(2)$};
  \node[flavor] (2) at (2,0) {\small $(A_1,A_3)$};
  \node[flavor] (3) at (-2,0) {\small $(A_1,A_3)$};
  \node[flavor] (4) at (0,2) {\small $(A_1,A_3)$};
\draw (3) -- (1) -- (2);
\draw (1) -- (4);
\end{tikzpicture}
\caption{The 4d $\mathcal{N}=2$ SCFT we consider in this paper is
 obtained by gauging the diagonal $SU(2)$ flavor symmetry of three
 copies of the $(A_1,A_3)$ theory. This theory is identical to the $(A_2,D_4)$ theory.}
\label{fig:quiver000}
\end{figure}

In this paper, we make a small but highly non-trivial generalization of this
4d/3d correspondence by identifying the first example of a 3d $\mathcal{N}=2$ CS matter theory
corresponding to a 4d $\mathcal{N}=2$ SCFT with a Coulomb branch
generator of integer $U(1)_r$ charge. Our focus
is on a particular 4d $\mathcal{N}=2$ SCFT obtained as follows. We first
take three copies of $(A_1,A_3)$ Argyres-Douglas (AD) theory \cite{Argyres:1995jj, Argyres:1995xn, Eguchi:1996vu,
Cecotti:2010fi, Xie:2012hs} and
consider gauging the diagonal $SU(2)$ flavor symmetry. The resulting
theory is described by the quiver diagram shown in
Fig.~\ref{fig:quiver000}. Since the $\beta$-function of the $SU(2)$
gauge coupling vanishes, this is a 4d $\mathcal{N}=2$ SCFT. In the
language of \cite{Cecotti:2010fi, Xie:2012hs}, this theory is called $(A_2,D_4)$ theory. Since the
Coulomb branch chiral ring of the $(A_2,D_4)$ theory contains one
generator of integer $U(1)_r$ charge, its R-twisted circle
compactification leads to a 3d $\mathcal{N}=4$ SCFT with a non-trivial
Coulomb branch. The aim of this paper is to identify a 3d
$\mathcal{N}=2$ CS matter theory that flows to this 3d $\mathcal{N}=4$
SCFT in the infrared.

Using the relation between the 4d Schur index and 3d half index, we identify the gauge group and matter
content of the 3d $\mathcal{N}=2$ CS matter theory. The only non-trivial
task is then to find an appropriate superpotential deformation that takes this CS
matter theory to the same IR fixed point as the R-twisted circle
reduction of $(A_2,D_4)$. We perform a careful analysis on expected
features of the IR fixed point and identify a monopole superpotential
that gives rise in the infrared to
\begin{itemize}
 \item a one-dimensional hyperk\"ahler Coulomb branch,
\item a superconformal R-charge consistent with a supersymmetry
      enhancement to $\mathcal{N}=4$~.
\end{itemize}
As a result, we find that the 3d Coulomb branch of the IR $\mathcal{N}=4$
SCFT is $\mathbb{C}^2/\mathbb{Z}_2$. The $SU(2)$ isometry of $\mathbb{C}^2/\mathbb{Z}_2$ implies that there is a 3d $\mathcal{N}=4$
flavor symmetry acting on the Coulomb branch. This $SU(2)$ symmetry has
no 4d counterpart and is an accidental symmetry in three dimensions.

Another striking consequence of our superpotential is that the Coulomb
branch R-symmetry is not proportional to the 4d $U(1)_r$ charge but
mixed with the accidental 3d flavor symmetry. A similar R-symmetry
mixing occurs in 3d reductions of 4d AD theories without R-twist
\cite{Buican:2015hsa}, and in those of 4d $\mathcal{N}\geq 3$ SCFTs
\cite{Nakanishi:2022fvr, Nakanishi:2024tnj}.

The rest of this paper is organized as follows. In Sec.~\ref{sec:A2D4},
we give a brief review of the $(A_2,D_4)$ theory and its Schur index. In
Sec.~\ref{sec:CS_A6}, we discuss a 3d $\mathcal{N}=2$ CS matter theory
that flows to the same IR $\mathcal{N}=4$ fixed point as the R-twisted
3d reduction of $(A_2,D_4)$. In particular, we identify a monopole
superpotential consistent with the one-dimensional Coulomb branch and
$\mathcal{N}=4$ supersymmetry expected in the infrared. In
Sec.~\ref{sec:summary}, we conclude with a discussion on a very
interesting future direction. In Appendix \ref{app:Macdonald}, we review how
to compute the Macdonald index of $(A_2,D_4)$. In Appendix \ref{app:sp},
we show the uniqueness of our monopole superpotential. In
Appendix.~\ref{app:w/o_r-twist}, we give a brief comparison of the Coulomb
branch of the IR $\mathcal{N}=4$ SCFT with that of the 3d reduction of
$(A_2,D_4)$ without an R-twist.

\section{$(A_2,D_4)$ theory}
\label{sec:A2D4}

In this section, we give a brief review of the $(A_2,D_4)$ theory and
its associated VOA. From the four-dimensional QFT viewpoint, the simplest
construction of  $(A_2,D_4)$  is as follows. We first take three copies of
the $H_1$ AD theory (or equivalently, $(A_1,A_3) = (A_1,D_3)$ theory or
$D_3(SU(2))$ theory) \cite{Argyres:1995jj, Argyres:1995xn,
Eguchi:1996vu}, and consider gauging the
diagonal $SU(2)$ flavor symmetry. Since the $\beta$-function of the
$SU(2)$ gauging vanishes, the resulting theory is an $\mathcal{N}=2$
SCFT. This theory is called the $\hat{\hat{E}}_6$ theory in
\cite{Cecotti:2011rv, Alim:2011ae, DelZotto:2015rca} and the ``$(3,2)$'' theory in \cite{Buican:2020moo}, and turns out to be
identical to the $(A_2,D_4)$ theory in the language of
\cite{Cecotti:2010fi, Xie:2012hs}.
 This theory also attracted attention as the
simplest example of 4d $\mathcal{N}=2$ non-Lagrangian SCFTs with $a=c$
\cite{Buican:2020moo, Kang:2021ccs, Kang:2022vab, Jiang:2024baj,
Kang:2024inx, Pan:2025gzh}. 

The $(A_2,D_4)$ theory has no $\mathcal{N}=2$ preserving flavor
symmetry, which implies that its associated VOA contains no affine
current. Moreover, one can show that the Higgs branch of $(A_2,D_4)$ is trivial, and
therefore the associated VOA is $C_2$-cofinite. Indeed, the relevant VOA
was conjectured in \cite{Buican:2020moo} to be a $C_2$-cofinite algebra
constructed in \cite{Feigin:2007sp, Feigin:2008}, as we will briefly review below.

In contrast to the trivial Higgs branch, the $(A_2,D_4)$ theory has a 
 Coulomb branch of complex dimension four, which is generated by
four Coulomb branch operators of dimensions
\begin{equation}
 \frac{4}{3}~,\quad \frac{4}{3}~,\quad \frac{4}{3}~,\quad 2~,
\label{eq:Coulomb}
\end{equation}
respectively. Here, the
three Coulomb branch operators of
dimension $4/3$ arise from the three copies of $H_1$, and one
dimension-two operator
arises from the $SU(2)$ vector multiplet. 
 Indeed, the presence of a dimension-two Coulomb branch operator implies
that $(A_2,D_4)$ has an exactly marginal coupling, which is identified
as the marginal $SU(2)$ gauge coupling.

\subsection{VOA associated with the $(A_2,D_4)$ theory}

The VOA associated in the sense of \cite{Beem:2013sza} with the
$(A_2,D_4)$ theory has been conjectured in \cite{Buican:2016arp} to be the so-called $\mathcal{A}(6)$
algebra studied in \cite{Feigin:2007sp, Feigin:2008}. It is generated by the
Virasoro stress tensor of dimension two, and a pair of fermionic Virasoro primaries of
dimension four. The Virasoro central charge is $c_\text{2d} = -24$,
which is required for the OPEs among these three generators to close. The
vacuum character of the $\mathcal{A}(6)$ algebra is evaluated in
\cite{Feigin:2007sp, Feigin:2008} as
\begin{equation}
 \mathcal{I}_{\mathcal{A}(6)}(q,z) \equiv
  \text{Tr}_{\mathcal{A}(6)}q^{L_0}z^{2f} =
  \frac{1}{(q)_\infty}\sum_{n=0}^\infty\sum_{j=-\frac{n}{2}}^{\frac{n}{2}}z^{2j}\left(q^{\frac{3n^2+5n}{2}}-q^{\frac{3n^2+7n+2}{2}}\right)~,
\end{equation}
where $(a)_n = \prod_{m=0}^{n-1}(1-a q^m)$ is the $q$-Pochhammer symbol and $f$ is the Cartan generator of $\mathfrak{sl}(2)$ under which the fermionic generators transform as a doublet. We normalize the character so
that the vacuum contributes one to it. It was then checked in
\cite{Buican:2020moo} that the Schur index of the $(A_2,D_4)$ theory coincides
with the above character with $z=-1$, i.e.,
\begin{equation}
 \mathcal{I}_{(A_2,D_4)}^\text{Schur}(q) = \mathcal{I}_{\mathcal{A}(6)}(q,-1)~,
\end{equation}
up to a very high order of $q$.

The above character of the $\mathcal{A}(6)$ algebra is known to have the following
fermionic formula \cite{Feigin:2007sp, Feigin:2008} (See Eq.~(4.69) of
\cite{Feigin:2008} in particular):
\begin{equation}
 \mathcal{I}_{\mathcal{A}(6)}(q,z) = \sum_{n_1,\cdots,n_7
  =0}^\infty z^{n_1-n_2}\frac{q^{\sum_{a,b=1}^7 \frac{1}{2}K_{ab}n_a
  n_b + \frac{1}{2}\sum_{a=1}^7 B_a n_a}}{\prod_{a=1}^7 (q)_{n_a}}~,
\end{equation}
where $K$ and $B$ are defined by
\begin{equation}
 K=\left(
\begin{array}{ccccccc}
 3 & 3 & 1 & 2 & 3 & 4 & 5\\
 3 & 3 & 1 & 2 & 3 & 4 & 5\\
 1 & 1 & 2 & 2 & 2 & 2 & 2\\
 2 & 2 & 2 & 4 & 4 & 4 & 4\\
 3 & 3 & 2 & 4 & 6 & 6 & 6\\
 4 & 4 & 2 & 4 & 6 & 8 & 8\\
 5 & 5 & 2 & 4 & 6 & 8 & 10\\
\end{array}
\right)~, \qquad B = \left(5,5,2,4,6,8,10\right)~.
\label{eq:KB}
\end{equation}
By setting $z=-1$, we find that the Schur index of the $(A_2,D_4)$
theory is expressed as
\begin{equation}
 \mathcal{I}^\text{Schur}_{(A_2,D_4)}(q) = \sum_{n_1,\cdots,n_7
  =0}^\infty \frac{q^{\sum_{a,b=1}^7 \frac{1}{2}K_{ab}n_a
  n_b }(-q^{\frac{1}{2}})^{\sum_{a=1}^7B_an_a}}{\prod_{a=1}^7
  (q)_{n_a}}~.
\label{eq:Schur_A6}
\end{equation}

\section{CS matter theory for the R-twisted reduction}
\label{sec:CS_A6}

In this section, we discuss the R-twisted 3d reduction of the
$(A_2,D_4)$ theory, and conjecture a 3d $\mathcal{N}=2$ CS matter
theory flowing into it in the infrared.
Since a
$2\pi$-rotation of $U(1)_r$ is introduced in the twisted reduction, all the Coulomb branch
operators with fractional $U(1)_r$ charges are lifted in the reduction and disappear from
the spectrum \cite{Dedushenko:2023cvd, Dedushenko:2018bpp,
ArabiArdehali:2024ysy, Gaiotto:2024ioj, ArabiArdehali:2024vli}. Since the $U(1)_r$ charge $r_\text{4d}$ of the
Coulomb branch operator is related to its dimension $\Delta_\text{4d}$
by 
\begin{equation}
 \Delta_\text{4d} = -r_\text{4d}~,
\end{equation}
this means that the three Coulomb branch operators of dimension $4/3$ in
Eq.~\eqref{eq:Coulomb} are removed by the R-twisted reduction.

In contrast to earlier
works, the 4d theory we are considering here includes a Coulomb branch
operator of {\it integer} $U(1)_r$ charge, which is not lifted in the
R-twisted reduction. Specifically, the dimension-two Coulomb
branch operator survives in the R-twisted reduction, and therefore we expect that the resulting 3d
$\mathcal{N}=4$ theory
is not a rank-zero theory but a theory with (quaternionic)
one-dimensional Coulomb branch. 

Below, we will identify a UV $\mathcal{N}=2$ CS matter theory that flows
to the same 3d $\mathcal{N}=4$ SCFT as the R-twisted 3d reduction of
$(A_2,D_4)$ described above.
In Sec.~\ref{subsec:Lagrangian}, we read off the gauge group and matter
content of the CS matter theory from an expression for the Schur
index. In Sec.~\ref{subsec:superpotential}, we discuss a monopole
superpotential that is consistent with the expected dimension of the
IR Coulomb branch. In Sec.~\ref{subsec:moduli_sp}, we show that this
superpotential leads to the IR Coulomb branch
$\mathbb{C}^2/\mathbb{Z}_2$. In Sec.~\ref{subsec:momentmap}, we read off
the superconformal R-charge and a flavor charge from expected features of flavor moment
maps. In Sec.~\ref{subsec:SCI}, we show that this superconformal
R-charge leads to an expression for the superconformal index consistent
with an enhanced $\mathcal{N}=4$ supersymmetry. In
Sec.~\ref{subsec:R-mix}, we show that the 3d Coulomb branch symmetry of
our $\mathcal{N}=4$ SCFT is not proportional to the $U(1)_r$ charge of
the parent 4d theory but mixed with an accidental flavor symmetry
acting on the 3d Coulomb branch. We also show the uniqueness of our
superpotential in Appendix \ref{app:sp}.

\subsection{3d Lagrangian from 4d Schur index}
\label{subsec:Lagrangian}

The matter content of the relevant CS matter theory can be identified by
recognizing \eqref{eq:Schur_A6} as the half index:
\begin{align}
I\!\!\!\!I (q)\equiv 
\mathrm{Tr} (-1)^{F} q^{j+\frac{R}{2}}
\end{align}
of the following 3d
$\mathcal{N}=2$ CS matter theory. Here $j$ is a rotation generator of the spacetime hemisphere
and $R$ is an R-charge. 
To see this, we first consider the $U(1)^7 = \prod_{a=1}^7 U(1)_a$
gauge group and identify $K$ as
the matrix of the mixed (gauge) Chern-Simons couplings. We also consider
a chiral multiplet of charge one for each $U(1)_a$ gauge group for
$a=1,\cdots,7$. Then the half index  of this CS matter theory with the
$(\mathcal{D},D_c) $ boundary condition \cite{Dimofte:2017tpi} is evaluated as
\begin{equation}
 I\!\!\!\!I_{(\mathcal{D},D_c)}(q) =
  \sum_{n_1,\cdots,n_7=0}^\infty\frac{q^{\sum_{a,b=1}^7
  \frac{1}{2}K_{ab}n_an_b}}{\prod_{a=1}^7(q)_{n_a}}\prod_{a=1}^7
  x_a^{-n_a}~,
\label{eq:half_A6}
\end{equation}
where $x_a$ are fugacities for the topological $U(1)^7$ symmetry.
The R-charge in this index computation is chosen to be
\begin{equation}
 R_H \equiv R_0 - \sum_{a=1}^7 B_a J_a~,
\label{eq:RH0}
\end{equation}
where $R_0$ is the R-charge that takes zero for the chiral multiplets,
and $J_a$ are the charges of the topological $U(1)^7$ symmetry
normalized so that the minimal non-vanishing value of $|J_a|$ is
one.
We now
see that the Schur index \eqref{eq:Schur_A6} is indeed identical to the half
index \eqref{eq:half_A6} under the condition
\begin{equation}
 x_a = -q^{-\frac{B_a}{2}}~,\qquad a = 1,2,\cdots,7~.
\label{eq:substitution}
\end{equation}

We interpret the above coincidence to mean that the $U(1)^7$ CS
matter theory described in the previous paragraph flows to the $\mathcal{N}=4$ SCFT
corresponding to the R-twisted 3d reduction of $(A_2,D_4)$ if an
appropriate superpotential deformation is turned on. We will identify
this appropriate superpotential deformation in Sec.~\ref{subsec:superpotential}.

Before identifying the correct superpotential, let us perform one
consistency check of the above interpretation here. If the above 
interpretation 
is correct, the
expression \eqref{eq:half_A6} with \eqref{eq:substitution} must have a special one-parameter
refinement corresponding to the Macdonald index of the 4d SCFT
\cite{Kim:2025klh}. Indeed, we experimentally find that
 \begin{equation}
 I\!\!\!\!I_{(\mathcal{D},D_c)}(q,t) =
  \sum_{n_1,\cdots,n_7=0}^\infty\frac{q^{\sum_{a,b=1}^7
  \frac{1}{2}K_{ab}n_an_b}}{\prod_{a=1}^7(q)_{n_a}}
  (-q^{\frac{1}{2}})^{\sum_{a=1}^7 B_a n_a}\left(\frac{t}{q}\right)^{\sum_{a=1}^7 v_a n_a}~,
\label{eq:refinement_A6}
\end{equation}
is identical to the Macdonald index of the $(A_2,D_4)$
theory (see appendix \ref{app:Macdonald}) up to high orders of $q$ and $t$, where $v_a$ are chosen as
\begin{equation}
 v^{(1)}_a \equiv (2,3,1,2,3,4,5)\qquad \text{or}\qquad v^{(2)}_a \equiv (3,2,1,2,3,4,5)~.
\label{eq:U1A}
\end{equation}
This reproduction of the Macdonald index suggests  that the 3d
$\mathcal{N}=2$ CS matter theory must preserve at least one
topological $U(1)$ symmetry corresponding to \eqref{eq:U1A}. Indeed, the 
factor $t/q$ is weighted by
\begin{equation}
 R_\text{4d} +r_\text{4d}
\label{eq:R-r}
\end{equation}
in the Macdonald index, which is preserved in the R-twisted 3d
reduction. 

We comment on a subtlety in the refinement of the half index when the Coulomb branch is nontrivial.
For a trivial Coulomb branch, i.e., a theory with no 3d $\mathcal{N}=4$ flavor symmetry acting on  the Coulomb branch, this refinement is written in \cite{Kim:2025klh} as  
\begin{align}
\mathrm{Tr}
(-1)^{ F}q^{j+\frac{R_H}{2}}
\left(\frac{t}{q} \right)^{\frac{R_H-R_C}{2}}. 
\label{eq:refine1}
\end{align}
Here, $R_H$ and $R_C$ are the Cartan generators of the 3d $\mathcal{N}=4$ R-symmetries acting on the Higgs and Coulomb branches, respectively. 
Then the relation between the UV charges and the IR R-charges can be read off from the refinement of the fermionic formula for the half index.
In contrast, when the Coulomb branch is nontrivial, a linear combination of topological charges that acts on the Coulomb branch as a flavor symmetry may enter the charge coupled to the fugacity $t/q$. As we will show in Sec.~\ref{subsec:R-mix}, in the twisted reduction, the 4d $U(1)$ R-charge is identified with a linear combination of a 3d R-charge and the Coulomb branch flavor charge $J^{\rm flavor}_{C}$. Consequently, the correct refinement corresponding to the Macdonald index \eqref{eq:refinement_A6} is not \eqref{eq:refine1}, but   
\begin{align}
I\!\!\!\!I_{(\mathcal{D},D_c)}(q,t)=\mathrm{Tr}
(-1)^{ F}q^{j+\frac{R_H}{2}}
\left(\frac{t}{q} \right)^{\frac{R_H-R_C \pm J^{\rm flavor }_C}{2}}\,,     
\end{align}
where the sign $\pm$ corresponds to two choices associated with
$v^{(i)}_a$, $i=1,2$.

\subsection{Monopole superpotential}
\label{subsec:superpotential}

Let us now consider which superpotential deformation takes the above $U(1)^7$
CS matter theory to the same $\mathcal{N}=4$ fixed point as the
R-twisted reduction of $(A_2,D_4)$. One constraint on the
superpotential is that it must preserve at least one $U(1)$ topological
symmetry corresponding to \eqref{eq:U1A}. There is indeed one
more constraint on the superpotential as discussed below.

Since the $(A_2,D_4)$ theory has no Higgs branch, so does its 3d
reduction. However, the 4d theory has a Coulomb
branch operator of integer $U(1)_r$ charge, which is not lifted by the R-twisted circle
compactification. Therefore, the Coulomb branch of the 3d
$\mathcal{N}=4$ SCFT is still non-trivial, in contrast to the R-twisted circle
compactifications of AD theories studied in the literature.
 Since we expect a 3d $\mathcal{N}=4$ superconformal symmetry in
the infrared, this 3d Coulomb branch is expected to be hyperk\"ahler. Then its complex dimension
must be even. This means that the Coulomb branch of the above 3d
$\mathcal{N}=4$ SCFT must have
complex dimension two, instead of dimension one.\footnote{
The fact that the complex dimension of the 3d Coulomb branch is twice
the number of 4d Coulomb branch 
generators
with integer 
$U(1)_r$ charge is a common feature of the 3d reductions of 4d $\mathcal{N}=2$ theories whose Coulomb branch operators all have integer scaling dimensions.
}

 Therefore, we need to find a
monopole superpotential of the UV CS-matter theory which leads to a
complex two-dimensional moduli space of vacua. This is the second
constraint on the superpotential.

We search for a monopole superpotential satisfying the above two
constraints. The superpotential must be composed of gauge-invariant chiral operators
of R-charge two. This is true for any R-charge, and therefore at least
true for $R_H$ that was used in the computation of the
half-index. Typical candidates for such chiral operators of $R_H=2$ are the following
gauge-invariant monopole operators:
\begin{equation}
 V_{\pmb{m}^{(k)}} \qquad (k=1,2, \cdots,7)~,
\label{eq:7monopoles}
\end{equation}
whose magnetic charges are given by
\begin{align}
 \pmb{m}^{(1)} &= (1,1,0,0,0,0,-1)~,
 \label{eq:magnetic1}
\\
\pmb{m}^{(2)} &= (0,0,2,-1,0,0,0)~,
 \label{eq:magnetic2}
\\
\pmb{m}^{(3)} &= (0,0,-1,2,-1,0,0)~,
 \label{eq:magnetic3}
\\
\pmb{m}^{(4)} &= (0,0,0,-1,2,-1,0)~,
 \label{eq:magnetic4}
\\
\pmb{m}^{(5)} &= (0,0,0,0,-1,2,-1)~,
 \label{eq:magnetic5}
\\
\pmb{m}^{(6)} &= (0,-1,1,-1,1,-1,1)~,
\\
\pmb{m}^{(7)} &= (-1,0,1,-1,1,-1,1)~.
\end{align}
There are more gauge-invariant chiral monopole operators of $R_H=2$,
which are obtained by multiplying the above seven with monopole
operators of $R_H=0$. We will discuss these other operators of $R_H=2$ later.

Given the above seven monopole operators, let us consider which of them can be included in the superpotential 
without breaking the preserved topological $U(1)$ symmetry
corresponding to \eqref{eq:U1A}. It is straightforward
to see that $V_{\pmb{m}^{(1)}},\,V_{\pmb{m}^{(2)}},\cdots,\,V_{\pmb{m}^{(5)}}$ are
neutral under both of
\eqref{eq:U1A}. In addition, $V_{\pmb{m}^{(6)}}$ preserves the topological
$U(1)$ corresponding to $v^{(1)}_a=(2,3,1,2,3,4,5)$ but breaks the one for
$v^{(2)}_a=(3,2,1,2,3,4,5)$, while $V_{\pmb{m}^{(7)}}$ preserves the latter but breaks
the former. Therefore, if we only want to preserve either one of
\eqref{eq:U1A}, one can at most include
$V_{\pmb{m}^{(1)}},\,V_{\pmb{m}^{(2)}},\,\cdots,\,V_{\pmb{m}^{(5)}}$
together with $V_{\pmb{m}^{(6)}}$ or $V_{\pmb{m}^{(7)}}$. 
However, based on the analysis presented below, we argue that the
correct choice for the superpotential is\footnote{If we only want to
preserve a $U(1)$ $\mathcal{N}=2$ flavor symmetry, we can include six
independent monopole operators in the superpotential. However, doing so
leads to a moduli space of complex one dimension, which is inconsistent with the fact that the IR Coulomb
branch is hyperk\"ahler. Thus, the hyperk\"ahler Coulomb branch implies
that the monopole superpotential must preserve a $U(1)^2$
$\mathcal{N}=2$ flavor symmetry.}
\begin{equation}
 W = \sum_{k=1}^5 V_{\pmb{m}^{(k)}}~.
\label{eq:superpot2}
\end{equation}
Note that including only five monopole operators in the superpotential means that two topological $U(1)$ symmetries are preserved. 
More precisely, one can show that the monopole superpotential given in \eqref{eq:superpot2} is the unique one that preserves precisely
the two topological $U(1)$ symmetries generated by $\sum_a v^{(i)}_a J_a$, $i=1,2$, and no other topological $U(1)$ symmetries; see Appendix~\ref{app:sp}.

One
linear combination of the topological $U(1)$ charges will be interpreted as \eqref{eq:R-r} associated with
$t/q$ in the Macdonald index, while another does not have a 4d origin;
it is an accidental global symmetry in three dimensions. The presence of
two $\mathcal{N}=2$ preserving global symmetry charges suggests that the
IR fixed point has an $\mathcal{N}=4$ ``flavor'' symmetry, i.e., a
global symmetry commuting with the $\mathcal{N}=4$ superconformal symmetry.

Recall that the 3d $\mathcal{N}=4$ SCFT has $\mathfrak{so}(4) \simeq
\mathfrak{su}(2)_C\times \mathfrak{su}(2)_H$ R-symmetry, where
$\mathfrak{su}(2)_C$ and $\mathfrak{su}(2)_H$ non-trivially act on the Coulomb and
Higgs branches, respectively. We denote by $R_C$ and $R_H$ the Cartans
of $\mathfrak{su}(2)_C$ and $\mathfrak{su}(2)_H$,
respectively.\footnote{We normalize them so that the smallest
non-vanishing value of $|R_C|$ and $|R_H|$ are both $1$.} Since
the Higgs branch is trivial (both in four and three dimensions) in our case, the
Higgs branch R-charge $R_H$
is naturally
identified as $R_\text{4d}$, up to the normalization. 
For the 3d half index to be identified as the 4d Schur index, the
normalization factor is chosen so that
\begin{equation}
 R_H = 2R_\text{4d}~.
\end{equation}
 On the other
hand, the Coulomb branch R-charge $R_C$ has two possible sources;
$r_\text{4d}$ from the parent 4d theory and the accidental ``flavor''
symmetry discussed in the previous paragraph. Indeed, since our Higgs
branch is trivial, the accidental $\mathcal{N}=4$ flavor symmetry must
have a non-trivial action on the 3d Coulomb branch. Thus, we expect that
\begin{equation}
 R_C = c_1 r_\text{4d} + c_2 J_C^\text{flavor}~,
\end{equation}
where $J_C^\text{flavor}$ is the charge for the accidental
$\mathcal{N}=4$ flavor symmetry, and $c_1$ and $c_2$ are real
constants such that $(c_1,c_2)\neq (0,0)$. We will show in Sec.~\ref{subsec:R-mix} that
$c_1=-2$ and $c_2=\pm 1$.

\subsection{Moduli space of vacua}
\label{subsec:moduli_sp}

The presence of the accidental flavor symmetry acting on the Coulomb
branch implies that the 3d Coulomb branch is
a hyperk\"ahler cone with at least a $U(1)$ isometry. Such a complex two-dimensional hyperk\"ahler cone
is always of the form
\begin{equation}
 \mathbb{C}^2/\mathbb{Z}_k
\end{equation}
for some positive integer $k$. Its isometry is just
$U(1)$ when $k\geq 3$, while it is enhanced to $SU(2)$ when $k=1,2$. 
Thus, we see that  our 3d Coulomb branch must be $\mathbb{C}^2/\mathbb{Z}_k$ for
some $k$. In other words, our CS matter theory must give rise in the infrared to Coulomb branch
operators that form the coordinate ring of
$\mathbb{C}^2/\mathbb{Z}_k$. In this sub-section, we show that this is
indeed the case.

To that end, we first search for monopole operators in the UV CS matter theory that
correspond to Coulomb branch operators of the IR fixed point. Since
every Coulomb branch operator is neutral under the Higgs
branch R-symmetry, they must have $R_H=0$. It
is straightforward to show that there is an infinite number of monopole
operators of $R_H=0$ in the UV CS matter theory. Indeed, we find that
bare monopole operators whose magnetic charge is of the following form
are all gauge-invariant and have $R_H=0$:
\begin{equation}
 \pmb{m}= n_1(0,-2,0,0,0,-1,2) + n_2(-2,0,0,0,0,-1,2) +
  n_3(-1,-1,0,0,0,-1,2)~,
\label{eq:mag}
\end{equation}
where $n_1,n_2$, and $n_3$ are non-negative integers.
Furthermore, we see that the chiral ring of these monopole operators is
generated by
\begin{equation}
 X\equiv V_{(0,-2,0,0,0,-1,2)}~,\qquad Y \equiv
  V_{(-2,0,0,0,0,-1,2)}~,\qquad Z \equiv V_{(-1,-1,0,0,0,-1,2)}
\label{eq:XYZ}
\end{equation}
subject to the operator relation\footnote{In general, the product of two monopole
operators of magnetic charge $m_1$ and $m_2$ gives rise to a monopole
operator of charge $m_1+m_2$ multiplied by a gauge-invariant
scalar operators arising from the matter sector \cite{Bullimore:2015lsa}. In our CS matter theory, there is no
such gauge-invariant scalar operator constructed out of the matter fields. }
\begin{equation}
 XY = Z^2~,
\end{equation}
which is precisely the ring relation for
$\mathbb{C}^2/\mathbb{Z}_2$. Therefore, identifying these monopole
operators with the IR Coulomb branch operators, we
conclude that the IR Coulomb branch is $\mathbb{C}^2/\mathbb{Z}_2$.

For the above identification to be correct, the operators $X$, $Y$, and $Z$ must be allowed to acquire nonzero VEVs without violating the F-term constraints arising from the superpotential \eqref{eq:superpot2}. To see that this is indeed the case at the semiclassical level, recall that a monopole operator with magnetic charge $\pmb{m}$ is written as
\begin{align}
V_{\bm{m}}\sim \exp\left({\bm m} \cdot {\bm \varphi}\right),
\end{align}
where ${\bm \varphi}=(\varphi_1,\ldots,\varphi_7)$ is a complex combination of the $U(1)^7$ vector multiplet scalars and the dual photons. 
At the semiclassical level,  the F-term constraints associated with the five monopole operators in the superpotential require
\begin{align}
{\bm m}^{(k)}\cdot {\bm \varphi}\longrightarrow-\infty,
\qquad k=1,\ldots,5.
\label{eq:Fconst}
\end{align}
Equivalently, the five linear combinations ${\bm m}^{(k)}\cdot {\bm\varphi}$ cannot remain finite and therefore should not be coordinates on the moduli space of vacua.
Since the five magnetic-charge vectors ${\bm m}^{(k)}$ are linearly independent, they remove five directions from the seven-dimensional space parameterized by the $\varphi_a$. 
Two finite directions remain. We may choose them to be
\begin{align}
\Phi_1\equiv (0,-2,0,0,0,-1,2)\cdot{\bm \varphi}, \quad 
\Phi_2\equiv (-2,0,0,0,0,-1,2)\cdot{\bm \varphi}.
\label{eq:gen1}
\end{align}
The coefficient vectors defining $\Phi_1$ and $\Phi_2$, together with the five vectors ${\bm m}^{(k)}$, form a basis of the space of linear combinations of the $\varphi_a$.  The corresponding monopole operators are  $X\sim e^{\Phi_1}$ and $Y\sim e^{\Phi_2}$, respectively. $Z\sim e^{(\Phi_1+\Phi_2)/2}$ also lies along these two finite directions.

Hence, when the UV CS matter theory is deformed by the
superpotential \eqref{eq:superpot2}, it gives rise to a set of monopole
operators corresponding to the Coulomb branch
$\mathbb{C}^2/\mathbb{Z}_2$. This orbifold has a non-trivial isometry,
which is consistent with the presence of the accidental ``flavor''
symmetry charge $J_C^\text{flavor}$. 
This is a non-trivial consistency check of
our superpotential \eqref{eq:superpot2}. 
In Appendix~\ref{app:w/o_r-twist}, we also give a comparison of this
$\mathbb{C}^2/\mathbb{Z}_2$ with the Coulomb branch of the 3d reduction
of $(A_2,D_4)$ without an R-twist.

\subsection{$SU(2)_C^\text{flavor}$ moment map and Coulomb branch R-symmetry}
\label{subsec:momentmap}

Our conclusion above is that the superpotential deformation
\eqref{eq:superpot2} takes the UV CS matter
theory to an $\mathcal{N}=4$ fixed point with
$\mathbb{C}^2/\mathbb{Z}_2$ as its Coulomb branch. Since its hyperk\"ahler isometry is
$SU(2)$, we expect that the accidental flavor symmetry associated with
$J_C^\text{flavor}$ is enhanced to $SU(2)$, which we denote by $SU(2)_C^\text{flavor}$.

For this to occur, there must be monopole operators that have
$R_{SC} = 1$ and belong to a triplet of
$SU(2)_C^\text{flavor}$, where
\begin{equation}
 R_{SC} \equiv \frac{R_H + R_C}{2}
\label{eq:RSCHC}
\end{equation}
is the superconformal R-charge of the IR $\mathcal{N}=4$
SCFT.\footnote{Recall that we have normalized $R_H$ and $R_C$ so that
the smallest non-vanishing values of $|R_H|$ and $|R_C|$ are both $1$.} These operators are flavor moment maps of $SU(2)_C^\text{flavor}$. Indeed, every
$\mathcal{N}=4$ flavor current multiplet contains a scalar chiral
operator of $R_{SC} = 1$.\footnote{Since the Coulomb branch symmetry
currents are neutral under $R_H$, they have $R_C=2$ and $R_{SC} = 1$.}
Below, we
show that there is indeed a choice of $R_{SC}$ under which $X,Y$, and $Z$
defined in \eqref{eq:XYZ} have the correct quantum numbers to be the
flavor $SU(2)_C^\text{flavor}$ moment maps.

To see this, recall first that the $U(1)$ transformation generated by any linear combination of $v^{(1)}_a$ and $v^{(2)}_a$ in \eqref{eq:U1A}
is preserved by our superpotential
\eqref{eq:superpot2}. We claim that the $U(1)$ symmetry generated by $\frac{R_C-R_H}{2}$ is identified as the average of
these two, i.e.,
\begin{equation}
 U(1)_{\frac{R_C-R_H}{2}}: \quad \frac{v^{(1)}_a+v^{(2)}_a}{2} = \left(\frac{5}{2},\frac{5}{2},1,2,3,4,5\right)~,
\label{eq:RCH}
\end{equation}
and  the $U(1)_C^\text{flavor}$ symmetry as the difference of
$v^{(1)}_a$ and $v^{(2)}_a$, i.e.,
\begin{equation}
U(1)_C^\text{flavor}:\quad v^{(2)}_a-v^{(1)}_a= (1,-1,0,0,0,0,0)~.
\label{eq:U1C}
\end{equation}
With these identifications, one can show that 
there is a set of monopole operators with the correct quantum numbers to be identified as the
$SU(2)_C^\text{flavor}$ moment maps. Indeed, the monopole operators
$X,Y$ and $Z$ defined in \eqref{eq:XYZ} have $\frac{R_C-R_H}{2} = 1$ under
\eqref{eq:RCH}. Since they are neutral under the Higgs branch R-symmetry, they
all have $R_H=0$. Thus, we see that $X,Y$ and $Z$ have
\begin{equation}
 R_{SC}= 1~.
\end{equation}
Furthermore, under \eqref{eq:U1C}, $X$ has charge $+2$, $Y$ has
charge $-2$, and $Z$ is neutral. These are
precisely the flavor charges of the $SU(2)_C^\text{flavor}$ moment maps. Thus, we have seen that \eqref{eq:RCH}
and \eqref{eq:U1C} are perfectly consistent with the observation that
$X,Y$ and $Z$ are flavor moment maps of $SU(2)_C^\text{flavor}$.
The fact that there is such a choice of $R_{SC}$, or equivalently $R_C -
R_H$, is further strong evidence for our choice of  superpotential \eqref{eq:superpot2}.

\subsection{Consistency with the superconformal index}
\label{subsec:SCI}

We now perform another consistency check of our identifications \eqref{eq:RCH} and
\eqref{eq:U1C} for $U(1)_{\frac{R_C-R_H}{2}}$ and $U(1)_C^\text{flavor}$ symmetries by computing the
superconformal index of the IR fixed point. Indeed, with the charge identifications in \eqref{eq:RCH} and \eqref{eq:U1C}, the
superconformal index 
\begin{align}
I_\text{SCI} (\mathfrak{q},T,x) &\equiv \text{Tr} (-1)^F\mathfrak{q}^{j +
 \frac{R_{SC}}{2}}T^{\frac{R_C-R_H}{2}}x^{J_C^\text{flavor}} 
\label{eq:SCI}
\end{align}
can be evaluated  by  the localization formula \cite{Kim:2009wb, Imamura:2011su} as
\begin{align}
I_\text{SCI} (\mathfrak{q},T,x)&=\sum_{m_1, \cdots, m_7 \in \mathbb{Z}}\oint \prod_{a=1}^7 \frac{dz_a}{2 \pi i z_a}  
\left( \prod_{a,b =1}^{7}  z_{a}^{K_{ab} m_b} \right)
(\mathfrak{q}^{-\frac{1}{2}}T)^{\frac{5}{2}(m_1+m_2) +\sum_{a=3}^7 (a-2) m_a } 
\nonumber \\
& \qquad \times x^{m_2-m_1}
\prod_{a=1}^7 (-\mathfrak{q}^{\frac{1}{2}} z_{a}^{-1})^{\frac{m_a +|m_a|}{2}} 
\frac{(z^{-1}_a \mathfrak{q}^{1+\frac{| m_a |}{2}} )_{\infty}}{(z_a \mathfrak{q}^{\frac{| m_a |}{2}} )_{\infty}} \nonumber 
\\[2mm]
&= 1 + \chi_{\bf 3}(x)T\mathfrak{q}^{\frac{1}{2}} + \left(-1 -
							  \chi_{\bf
							  3}(x)+T^2
							  \chi_{\bf
							  5}(x)
							 \right)\mathfrak{q}
\nonumber\\
&\qquad  + \left(T + \frac{1}{T} -T\chi_{\bf 5}(x) +T^3\chi_{\bf
 7}(x)\right)\mathfrak{q}^{\frac{3}{2}} + \mathcal{O}(\mathfrak{q}^2)~.
\end{align}
Here $\chi_{\bm{2k+1}}(x)$ is the character of the spin-$k$ representation of $SU(2)$,
$j$ is the spin, $\mathfrak{q}$ is the superconformal fugacity, $T$ is the
fugacity for $U(1)_{\frac{R_C-R_H}{2}}$, and $x$ is the fugacity for the $SU(2)$ Coulomb
branch symmetry.\footnote{Note here that \eqref{eq:RSCHC} implies
$R_{SC} = R_H +\frac{R_C-R_H}{2}$. Combining this with \eqref{eq:RH0},
we find  $R_\text{SC}= R_0 - \frac{1}{2}\sum_{a=1}^7 B_a J_a$, where
$B_a=(5,5,2,4,6,8,10)$ as shown in \eqref{eq:KB}, and $R_0$ is the
R-charge so that the matter chiral multiplets have $R_0=0$.} We see that
there are a term $-\mathfrak{q}$ corresponding to the
$U(1)_{\frac{R_C-R_H}{2}}$ current multiplet (included in the
$\mathcal{N}=4$ stress-tensor multiplet) and terms
$\mathfrak{q}^{\frac{3}{2}}(T+\frac{1}{T})$ corresponding to the extra
SUSY current multiplets.\footnote{One can use $\text{Tr} (-1)^{R_\text{SC}}\mathfrak{q}^{j +
 \frac{R_{SC}}{2}}T^{\frac{R_C-R_H}{2}}x^{J_C^\text{flavor}}$ instead of
 \eqref{eq:SCI} as the
 definition of the superconformal index, in which case the extra SUSY
 multiplets contribute
 $-\mathfrak{q}^{\frac{3}{2}}(T+\frac{1}{T})$. These two definitions are
 related to each other by $\mathfrak{q}^{\frac{1}{2}} \to -\mathfrak{q}^{\frac{1}{2}}$.}

Furthermore, we see that in the Higgs branch limit ($\mathfrak{q}\to 0$ with 
$T\mathfrak{q}^{-\frac{1}{2}}$ fixed) which evaluates the Hilbert series of 
Higgs branch vacua,  the index reduces to one, i.e.,
\begin{equation}
 \lim_{\mathfrak{q}\to 0  \atop T \mathfrak{q}^{-1/2} ;\text{fixed}} \!\!\!  I_{\rm SCI} = 1~.
\end{equation}
This is consistent with the fact that the IR $\mathcal{N}=4$ fixed point
has a trivial Higgs branch.
On the other hand, in the Coulomb branch limit ($\mathfrak{q}\to 0$ with
$\mathfrak{t} \equiv T\mathfrak{q}^{\frac{1}{2}}$ fixed), which evaluates the Hilbert series of Coulomb branch vacua,
we find
\begin{equation}
 \lim_{\mathfrak{q}\to 0 \atop \mathfrak{t}  =T \mathfrak{q}^{1/2} ; \text{fixed}}
\!\! \! I_{\rm SCI}= 1 + \chi_{\bf 3} (x) \mathfrak{t} + \chi_{\bf 5}(x)
  \mathfrak{t}^2 + \chi_{\bf 7}(x) \mathfrak{t}^3 + \cdots\,,
\end{equation}
This can be identified as the first four terms of the
$\mathfrak{t}$-expansion of the Hilbert series of $\mathbb{C}^2/\mathbb{Z}_2$:
\begin{equation}
 I_{\mathbb{C}^2/\mathbb{Z}_2}(\mathfrak{t},x) = \frac{1-\mathfrak{t}^2}{(1-\mathfrak{t})(1-\mathfrak{t}x^2)(1-\mathfrak{t}x^{-2})} = \sum_{k=0}^\infty \chi_{\bm{2k+1}}(x)\mathfrak{t}^k~.
\end{equation}
This is consistent with our observation that the Coulomb branch of the
IR fixed point is $\mathbb{C}^2/\mathbb{Z}_2$.

All the above results provide strong evidence for our identifications
\eqref{eq:RCH}  and \eqref{eq:U1C} of
$\frac{R_C-R_H}{2}$ and $J_C^\text{flavor}$ symmetry, and the
presence of such a special choice of $\frac{R_C-R_H}{2}$ and
$J_C^\text{flavor}$ strongly suggests that our monopole superpotential
\eqref{eq:superpot2} correctly takes the UV CS matter theory to an
$\mathcal{N}=4$ fixed point.

\subsection{R-symmetry mixing}
\label{subsec:R-mix}

The identification \eqref{eq:RCH} implies that the 3d Coulomb branch
symmetry $R_C$ is not proportional to the 4d $U(1)_r$ charge
$r_\text{4d}$, but is rather a linear combination of the 4d $U(1)_r$
charge and the 3d accidental global symmetry charge $J_C^\text{flavor}$.

To see this, first recall  that the
Macdonald index is defined by
\begin{equation}
 \text{Tr} (-1)^F
  q^{R_\text{4d}+j_1+j_2}\left(\frac{t}{q}\right)^{R_\text{4d}+r_\text{4d}}~,
\label{eq:Macdonald}
\end{equation}
where $(j_1,j_2)$ is the spin, $R_\text{4d}$ is the $SU(2)_R$
charge, and $r_\text{4d}$ is the $U(1)_r$ charge of local operators.
Since \eqref{eq:refinement_A6} coincides with the Macdonald
index of the $(A_2,D_4)$ theory, we identify
\begin{equation}
R_\text{4d} + r_\text{4d}~,
\end{equation}
as corresponding to (either one of) \eqref{eq:U1A} in three dimensions.
In addition, we identify 
\begin{equation}
 R_H= 2R_\text{4d}~,\qquad j= j_1 + j_2
\label{eq:RH}
\end{equation}
so that \eqref{eq:Macdonald} can be identified as the refined half-index
$\text{Tr}(-1)^F q^{j + \frac{R_H}{2}} \left(\frac{t}{q}\right)^{v_a J_a}$,
where $J_a$ are the topological charges.
This implies that
\begin{equation}
 \frac{R_H}{2} + r_\text{4d} = \left\{
\begin{array}{l} 
-2J_1 - 3J_2 -J_3-2J_4-3J_5-4J_6 - 5J_7
\\[2mm]
-3J_1-2J_2-J_3-2J_4-3J_5-4J_6-5J_7\\
\end{array}
\right.~,
\label{eq:RH2}
\end{equation}
depending on which of \eqref{eq:U1A} we choose.\footnote{Here, the minus
signs on the right-hand side come from the fact that
$(-n_1,-n_2,\cdots,-n_7)$ is the topological charges in the convention
that each of the seven chiral multiplets in the CS matter theory has an electric
charge $+1$.}
On the other hand,  the $U(1)_{\frac{R_C-R_H}{2}}$ was identified in 
\eqref{eq:RCH} as, i.e.,
\begin{equation}
 \frac{R_C-R_H}{2} = \frac{5}{2}J_1 + \frac{5}{2}J_2 +
  J_3+2J_4+3J_5+4J_6+5J_7~.
\label{eq:RCH2}
\end{equation}
Combining these relations, we find $-r_\text{4d} - \frac{R_C}{2} =
\pm \frac{1}{2}(J_1-J_2) = \pm \frac{1}{2}J_C^\text{flavor}$, or
equivalently
\begin{equation}
 R_C = -2r_\text{4d} \mp J_C^\text{flavor}~.
\label{eq:RC}
\end{equation}
The presence of two options for the sign in front of $J_C^\text{flavor}$ is natural,
since the Weyl group of $SU(2)_C^\text{flavor}$ induces
$J_C^\text{flavor}\to -J_C^\text{flavor}$. Indeed, this Weyl group can
be regarded as a reason for two possibilities of \eqref{eq:U1A} in the
identification of the 4d Macdonald index as a 3d refined half index.

Let us briefly discuss what \eqref{eq:RC} implies. In four dimensions, we started with a 4d $\mathcal{N}=2$
SCFT with $\mathfrak{su}(2)_R \times \mathfrak{u}(1)_r$ global symmetry. In
the R-twisted 3d reduction, this theory is expected to flow to a 3d
$\mathcal{N}=4$ SCFT with $\mathfrak{so}(4)_R\simeq
\mathfrak{su}(2)_H\times \mathfrak{su}(2)_C$ R-symmetry. In addition, we know
that our 3d theory has a flavor $\mathfrak{su}(2)^\text{flavor}_C$ symmetry acting on the 3d
Coulomb branch. Since it  does not exist in the
parent 4d theory, this $\mathfrak{su}(2)_C^\text{flavor}$ symmetry is an accidental symmetry in three
dimensions. In total, the 3d theory has $\mathfrak{su}(2)_H\times
\mathfrak{su}(2)_C\times \mathfrak{su}(2)_C^\text{flavor}$. What we have
seen in \eqref{eq:RH}, \eqref{eq:RH2} and \eqref{eq:RC} is that the 3d $\mathfrak{su}(2)_H$ symmetry is naturally
identified as the 4d $\mathfrak{su}(2)_R$ symmetry, but the (Cartan
sub-algebra of) 3d
$\mathfrak{su}(2)_C$ symmetry is {\it not} identified as
the 4d $\mathfrak{u}(1)_r$. Instead, the 4d $\mathfrak{u}(1)_r$ is
identified as a linear combination of the Cartans
of $\mathfrak{su}(2)_C$ and $\mathfrak{su}(2)_C^\text{flavor}$. In other words, the 3d (Coulomb
branch) R-symmetry charge $R_C$ is a linear combination of the 4d
$U(1)_r$ charge and an accidental Coulomb branch symmetry in three
dimensions. A similar R-symmetry mixing was observed in a related but
different 3d reduction of AD theories \cite{Buican:2015hsa}.

\section{Summary and discussions}
\label{sec:summary}

In this paper, we have studied the R-twisted 3d reduction of the
$(A_2,D_4)$ theory, and conjectured a 3d $\mathcal{N}=2$ CS matter theory
flowing to the same IR fixed point as this R-twisted 3d reduction. A
crucial difference from related works in the literature is that our 4d
theory has a Coulomb branch generator of integer $U(1)_r$ charge, and
therefore the R-twisted 3d reduction does not completely get rid of the Coulomb
branch operators from the spectrum. As a result, the resulting 3d
$\mathcal{N}=4$ SCFT has a non-trivial Coulomb branch. From the
knowledge of the Schur/Macdonald index in four dimensions, we have identified the
matter content of the $\mathcal{N}=2$ CS matter theory. Then we have
carefully studied the spectrum of monopole operators of the theory and
conjectured a monopole superpotential that gives rise to the following:
\begin{itemize}
 \item a non-trivial hyperk\"ahler Coulomb branch as expected,
 \item a superconformal R-charge consistent with $\mathcal{N}=4$ SUSY,
 \item an expression for the superconformal index consistent with the IR $\mathcal{N}=4$ SUSY enhancement
       and expected features of the Coulomb/Higgs branches.
\end{itemize}

As a result, we have found that the Coulomb branch of the IR $\mathcal{N}=4$
fixed point is $\mathbb{C}^2/\mathbb{Z}_2$, and therefore the IR theory
has an $\mathcal{N}=4$ $SU(2)$ flavor symmetry acting on the Coulomb
branch. 
This $SU(2)$ flavor symmetry is purely three-dimensional, and has no 4d
origin. 
 Moreover, the charge $J_C^\text{flavor}$ of this
accidental flavor symmetry is combined with the 4d $U(1)_r$ charge
$r_\text{4d}$ to realize (the Cartan of) the Coulomb branch
R-symmetry $\mathfrak{su}(2)_C\subset \mathfrak{so}(4)_R$. This kind of
R-symmetry mixing between the 4d $U(1)_r$ and a 3d topological symmetry
has been observed in different contexts \cite{Buican:2015hsa, Nakanishi:2022fvr, Nakanishi:2024tnj}, but is new in the study of
R-twisted 3d reductions of 4d $\mathcal{N}=2$ SCFTs.

A very interesting future direction is to perform a similar analysis on
the R-twisted reductions of 4d $\mathcal{N}=2$ Lagrangian SCFTs and
isolated SCFTs obtained at cusps of conformal manifolds of Lagrangian SCFTs. When
the 4d theory has a ($\mathcal{N}=2$ preserving) Lagrangian description,
its Coulomb branch chiral ring always contains a generator of integer
$U(1)_r$ charge. Therefore, an accidental 3d $\mathcal{N}=4$ flavor
symmetry acting on the 3d Coulomb branch could occur as in our
case. Taking into account this possibility would be extremely important in identifying the correct
superpotential for the 3d $\mathcal{N}=2$ CS matter theory. Among other
theories, it would be interesting to identify 3d $\mathcal{N}=2$ CS
matter theories (with appropriate superpotentials) that flow to the same
IR fixed points as the R-twisted compactifications of 4d $\mathcal{N}=2$
$SU(2)$ superconformal QCD \cite{Seiberg:1994aj} and Minahan--Nemeschansky $E_6, E_7$, and $E_8$
theories \cite{Minahan:1996fg, Minahan:1996cj}. Such a generalization will lead to the discovery
of 3d $\mathcal{N}=2$ CS matter theories whose IR $\mathcal{N}=4$ fixed
points contain $D_4, E_6, E_7$ and $E_8$ flavor symmetries acting on their
Higgs branches, respectively. If the IR theory also has
a flavor symmetry acting on the 3d Coulomb branch, the full $\mathcal{N}=4$
flavor symmetry algebra is a larger one that respectively contains $D_4,E_6,E_7$ and $E_8$ as a sub-algebra.

\section*{Acknowledgements}

We are grateful to Matthew Buican, Mykola Dedushenko, Dongmin Gang, Sungjoon Kim and
Pietro Longhi for helpful discussions.
In particular, T.~N. thanks Matthew Buican for illuminating
discussions in a separate but related collaboration. 
T.~N.'s research is partially supported by JSPS KAKENHI Grant
Numbers JP21H04993, JP23K03394, JP23K03393 and JP26H01997. 
S.~T.'s research is supported by JST SPRING, Grant
Number JPMJSP2139.

\appendix

\section{Macdonald index of $(A_2,D_4)$}
\label{app:Macdonald}

Here, we give a brief review of how to evaluate the Macdonald index of the $(A_2,D_4)$
theory. Since the $(A_2,D_4)$ theory has a weakly coupled description
shown in Fig.~\ref{fig:quiver000}, one can evaluate the index as
\begin{equation} 
\mathcal{I}_\text{Mac}(q,t) = \oint_{|z|=1} \frac{dz}{2\pi i z}\,
 \Delta(z)\,
 \mathcal{I}_\text{vec}(q,t,z)\big(\mathcal{I}_{(A_1,A_3)}(q,t,z)\big)^3~,
\label{eq:Mac}
\end{equation}
where $\Delta(z)\equiv \frac{1}{2}(1-z^2)(1-z^{-2})$ comes from the
Haar measure of $SU(2)$, and $\mathcal{I}_\text{vec}(q,t,z) \equiv
P.E.\left[\frac{-t-q}{1-q}(z^2+1+z^{-2})\right]$ is the Macdonald
index of an $SU(2)$ vector multiplet. The last factor
$\mathcal{I}_{(A_1,A_3)}(q,t,z)$ is the Macdonald index of an $(A_1,A_3)$
theory, and a formula to compute it was conjectured in
\cite{Buican:2015tda}. Plugging the result of this formula into
\eqref{eq:Mac}, one can evaluate $\mathcal{I}_\text{Mac}(q,t)$ as a
series in powers of $q$ and $t$.

\section{Uniqueness of the monopole superpotential}
\label{app:sp}
We show that the only gauge-invariant monopole operators with R-charge $R_H=2$ that preserve the $U(1)^2$ symmetry generated by $v^{(i)}_a$, $i=1,2$ are $V_{{\bm m}^{(k)}}$, $k=1,2,3,4,5$.

Let us consider a general gauge-invariant BPS dressed monopole operator with $R_H=2$:
\begin{align}
\left(\prod_{a=1}^7 \phi_a^{n_a}\right) V_{\bm m}\,.
\end{align}
Here $\phi_a$, $a=1,\cdots, 7$ are the scalars in the seven chiral multiplets.
Note that ${\bm m} \equiv (m_1,\cdots,m_7)\in\mathbb{Z}^7$ and
${\bm n} \equiv (n_1,\cdots,n_7)\in\mathbb{Z}_{\geq 0}^7$ satisfy  $n_a m_a=0$ for $a=1,\cdots,7$.

The conditions for preserving the $U(1)^2$ symmetry generated by $v^{(i)}_a$, $i=1,2$, are
\begin{align}
\sum_{a=1}^7 v^{(1)}_a m_a=0 \quad  \text{and} \quad 
\sum_{a=1}^7 v^{(2)}_a m_a=0\,.
\end{align}
Taking the difference between these two conditions gives
\begin{align}
m_1=m_2\,.
\label{eq:m1m2}
\end{align}
Adding the two conditions gives
\begin{align}
5(m_1+m_2)+2(m_3+2m_4+3m_5+4m_6+5m_7)=0\,.
\label{eq:average}
\end{align}
Together with the condition \eqref{eq:m1m2}, the $R_H=2$ condition:
\begin{align}
\sum_{a=1}^{7}p_a=2,
\end{align}
where we defined $p_a\equiv {\rm max}(m_a,0)$, implies that the positive part of the magnetic charge must be
\begin{align}
(p_1,\cdots,p_7)=(1,1,0,\cdots,0)
 \quad \text{or} \quad (0,\cdots,0, 2, 0,\cdots,0)\,.
\label{eq:pcond}
\end{align}

Next, recall that the gauge-invariance condition for the dressed monopole operator is given by
\begin{align}
\sum_{a=1}^7 K_{ab}m_a+n_b=p_b\,,
\label{eq:gaugeinv}
\end{align}
where $K_{ab}$ is the CS level given in \eqref{eq:KB}.
Multiplying \eqref{eq:gaugeinv} by ${\bm m}^T$ from the left, we obtain
\begin{align}
{\bm m}^{T}K{\bm m}=\sum_{a=1}^{7}p_a^2
\end{align}
Then it  follows from \eqref{eq:pcond} that
\begin{align}
{\bm m}^{T}K{\bm m}=2 \quad \text{or} \quad 4\,.
\end{align}
On the other hand, any $(m_1,\cdots,m_7)$ satisfying \eqref{eq:m1m2} and
\eqref{eq:average} can be uniquely expressed as
\begin{align}
{\bm m}=\sum_{i=1}^{5}c_i{\bm m}^{(i)}, \quad c_i\in\mathbb{Z}\,.
\end{align}
The quadratic form ${\bm m}^{T}K{\bm m}$ can then be written as
\begin{align}
{\bm m}^{T}K{\bm m}=2\left[c_1^2+f(c_2,c_3,c_4,c_5)\right]
\label{eq:cquadra}
\end{align}
Here we defined 
\begin{align}
f(c_2,c_3,c_4,c_5)
\equiv c_2^2+(c_2-c_3)^2+(c_3-c_4)^2+(c_4-c_5)^2+c_5^2.
\label{eq:Edef}
\end{align}
The function $f$ is the sum of the squares of the successive differences of the integer sequence
\begin{align}
0,\ c_2,\ c_3,\ c_4,\ c_5,\ 0.
\end{align}
Therefore, if this sequence is nonzero, there must be at least one jump away from zero and at least one jump back to zero. It follows that
\begin{align}
f=0\qquad\text{or}\qquad f\geq2.
\label{eq:Ebound}
\end{align}
Moreover, $f=2$ if and only if $(c_2,c_3,c_4,c_5)$ is equal to $+1$ or $-1$ on a consecutive interval and vanishes outside that interval.

First we consider the case ${\bm m}^{T}K{\bm m}=2$.  The solutions to \eqref{eq:cquadra}  are
\begin{align}
c_1=\pm1\,, \quad c_2=c_3=c_4=c_5=0\,.
\end{align}
However, $c_1=-1$ does not satisfy the condition $R_H=2$. For $c_1=1$, we  obtain the magnetic charge
${\bm m}=(1,1,0,0,0,0,-1)$, which is ${\bm m}^{(1)}$ in \eqref{eq:magnetic1}.

Next, consider the case ${\bm m}^{T}K{\bm m}=4$,  In this case, we have
$c_1^2+f=2$.
Since $f$ is either zero or greater than or equal to two, and in particular cannot be equal to one, we must have
\begin{align}
c_1=0,\qquad f=2.
\end{align}
Therefore, $(c_2,c_3,c_4,c_5)$ is equal to $+1$ or $-1$ on a consecutive interval and vanishes outside that interval.

First, suppose that the value on the interval is $+1$ and that the interval consists of a single component. Then,
\begin{align}
(c_2,c_3,c_4,c_5)
=(1,0,0,0),\quad
(0,1,0,0),\quad
(0,0,1,0),\quad
(0,0,0,1).
\end{align}
These four choices give ${\bm m}^{(2)},  {\bm m}^{(3)},  {\bm m}^{(4)},  {\bm m}^{(5)}$ in \eqref{eq:magnetic2}-\eqref{eq:magnetic5}, respectively. 

On the other hand, if the interval on which the entries are equal to $+1$ has length greater than or equal to two, the positive part of the magnetic charge consists of two components equal to one. Therefore, $\sum_a p_a^2=2$.
However, this contradicts ${\bm m}^{T}K{\bm m}=\sum_a p_a^2=4$. Hence, these possibilities are excluded.

Next, consider the case in which the entries on the interval are equal to $-1$. If the interval starts at $c_2$, the sum of the positive components of the magnetic charge is one, and hence the condition $R_H=2$ is not satisfied. In all the other cases, the positive part of the magnetic charge consists of two components equal to one, so that
$ \sum_a p_a^2=2$. This again contradicts ${\bm m}^{T}K{\bm m}=4$.

Finally, these five magnetic charges: ${\bm m}^{(1)}, \cdots {\bm m}^{(5)}$ satisfy 
\begin{align}
\sum_{a=1}^7 K_{ab}m_a={\rm max}(m_a, 0)\,,
\end{align}
which means the dressed factor ${\bm n}=0$.
Thus, all the gauge-invariant  monopole operators satisfying the above conditions are the bare monopole operators: $V_{{\bm m}^{(1)}},
V_{{\bm m}^{(2)}}, V_{{\bm m}^{(3)}}, V_{{\bm m}^{(4)}}, V_{{\bm m}^{(5)}}$.
Note that if the superpotential contained fewer of these five monopole operators,  additional topological $U(1)$ symmetries would remain unbroken.

\section{Relation to 3d reduction without R-twist}
\label{app:w/o_r-twist}

Here, we compare the Coulomb branch of the R-twisted 3d reduction of
$(A_2,D_4)$ that we have identified in the main text, with that of its
direct 3d reduction without an R-twist.

\begin{figure}
\centering
 \begin{tikzpicture}[vec/.style={thick,-latex,draw=black},gauge/.style={circle,draw=black,inner sep=0pt,minimum size=12mm},flavor/.style={rectangle,draw=black,inner sep=0pt,minimum size=10mm},auto]
  \node[gauge] (1) at (0,0) {\small $SU(2)$};
  \node[gauge] (2) at (2.3,0) {\small $U(1)$};
  \node[gauge] (3) at (-2.3,0) {\small $U(1)$};
  \node[gauge] (4) at (0,2.3) {\small $U(1)$};
\draw (3) -- (1) -- (2);
\draw (1) -- (4);
\end{tikzpicture}
\caption{The quiver diagram describing the 3d reduction of the $(A_2,D_4)$
 theory without R-twisting. Each circle with $G$ inside stands for an
 $\mathcal{N}=4$ vector multiplet of gauge group $G$, and each line segment stands for a bifundamental
 hypermultiplet.}
\label{fig:quiver1}
\end{figure}
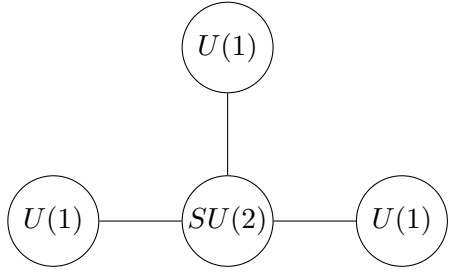

To that end, let us consider  a weak marginal coupling limit of the
$(A_2,D_4)$ theory, which involves three copies of $(A_1,A_3)$ theories
coupled to an $SU(2)$ vector multiplet; each $(A_1,A_3)$ has an $SU(2)$
flavor symmetry and the vector multiplet is gauging the diagonal
$SU(2)$ flavor symmetry. When we compactify this theory on $S^1$ without R-twisting, each
$(A_1,A_3)$ sector flows to the 3d $\mathcal{N}=4$ $U(1)$ gauge theory
coupled to two hypermultiplets of charge one
\cite{Xie:2012hs}.\footnote{To be precise, the author of
\cite{Xie:2012hs} identified this $U(1)$ gauge theory as the 3d mirror
theory of $(A_1,A_3)$. Since this $U(1)$ gauge theory is self-mirror,
one can also regard it as the direct 3d reduction of $(A_1,A_3)$.} This theory has an $SU(2)$
Higgs branch symmetry, which is naturally identified as the 4d flavor
$SU(2)$ symmetry of $(A_1,A_3)$. Let us take three copies of
this $U(1)$ gauge theory and consider the diagonal $SU(2)$ gauging of the
Higgs branch symmetry, by coupling them with a 3d $\mathcal{N}=4$ $SU(2)$
vector multiplet. The resulting $\mathcal{N}=4$ theory is described by a quiver
diagram shown in Fig.~\ref{fig:quiver1}.\footnote{We thank Matthew
Buican very much for pointing out that this quiver gauge theory captures the direct
3d reduction of the $(A_2,D_4)$ theory without an R-symmetry twist.}

We expect that this 3d $\mathcal{N}=4$ gauge theory flows to an
$\mathcal{N}=4$ SCFT in the infrared, which we identify as the direct 3d
reduction of $(A_2,D_4)$ without an R-twist. This is clearly not the same theory
as the 3d $\mathcal{N}=4$ SCFT obtained by the R-twisted circle
compactification of $(A_2,D_4)$. Indeed, in the latter case, 4d Coulomb
branch operators of fractional $U(1)_r$ charges are all removed from the
3d Coulomb branch chiral ring, while in the former case they are not. As a result, the 3d theory shown
in Fig.~\ref{fig:quiver1} has a Coulomb branch of quaternionic dimension four.

Despite the different dimension of the Coulomb branch, we expect that
the 3d $\mathcal{N}=4$ theory shown in Fig.~\ref{fig:quiver1} still
captures a part of the Coulomb branch physics of the 3d $\mathcal{N}=4$
SCFT that we have studied in the main text. In particular,
we expect that an appropriate subspace of the four-dimensional Coulomb branch
 of the former theory is identified as
the Coulomb branch of the latter.

Indeed, the Coulomb branch of the theory of Fig.~\ref{fig:quiver1} is
argued to be
$\mathbb{C}^8/\mathbb{Z}_2$ on which an $USp(8)$ flavor symmetry acts
\cite[the end of Sec.~5]{Gaiotto:2008ak}. It is therefore natural to
identify the Coulomb branch $\mathbb{C}^2/\mathbb{Z}_2$ discussed in
Sec.~\ref{subsec:moduli_sp} as a subspace of this
$\mathbb{C}^8/\mathbb{Z}_2$. It would be interesting to identify
generators of the Coulomb branch chiral sub-ring corresponding to this
subspace explicitly in terms of monopole operators of the theory of
Fig.~\ref{fig:quiver1}, which we leave for future work.

\bibliography{VOA}

\end{document}